\documentclass[12pt]{article}
\usepackage{latexsym}
\usepackage{epsfig}

\def\bg#1{\mbox{\boldmath$#1$}}

\newcommand{\Det}{\mbox{det}}

\newcommand{\del}{\partial}
\newcommand{\beq}{\begin{eqnarray}}
\newcommand{\eeq}{\end{eqnarray}}
\newcommand{\be}{\begin{eqnarray*}}
\newcommand{\ee}{\end{eqnarray*}}
\newcommand{\bk}{{\bf k}}
\newcommand{\bp}{{\bf p}}

\newcommand{\bx}{{\bf x}}

\newcommand{\ra}{\rightarrow}

\newcommand{\ve}{\varepsilon}

\newcommand{\om}{{\omega}}

\begin{document}

\centerline{\Large\bf{Symmetric and conserved energy-momentum }}
\bigskip
\centerline{\Large\bf{tensors in moving media\footnote{Written version of talk presented at {\it Scalars 2011}, Warsaw, August 26-29, 2011.}}}
\bigskip
\centerline{Finn Ravndal}
\bigskip
\centerline{\it Department of Physics, University of Oslo, Blindern, 0316 Oslo, Norway.}

\begin{abstract}

\small{A symmetric and conserved energy-momentum tensor for a scalar field in a moving medium is derived using the Gordon metric. When applied to an electromagnetic field, the method gives a similar result. This approach thus points a way out of the old Abraham-Minkowski controversy  about the correct energy-momentum tensors  for the electromagnetic field in a material medium. The new tensor describes the properties of the field while the Abraham tensor contains information about the field coupled to the medium.}

\end{abstract}

In a medium with refractive index $n$  the velocity of light is reduced by the same factor. This comes in conflict with the ordinary Lorentz transformations which involve the velocity of light in vacuum. It is therefore not {\it a priori} clear how a covariant description can be established.  Already Minkowski in his first unification of electric and magnetic fields in a four-dimensional spacetime considered this problem\cite{Minkowski}. He derived an electromagnetic energy-momentum tensor that was not symmetric and conserved in only one index. 

Symmetry of the energy-momentum tensor is needed for the conservation of the total angular momentum of the system. Abraham thus proposed shortly afterwards a symmetric tensor, but no longer conserved\cite{Abraham}. It should be accompanied by a corresponding volume force acting on the medium.  After the development of the general theory of relativity, Gordon pointed out that the reduced speed of light in a medium was mathematically equivalent to considering the electromagnetic field in a non-trivial metric\cite{Gordon}. By letting this metric depend on position, he  could then derive the energy-momentum tensor by standard methods. This tensor is the source of the gravitational field described by the metric and will therefore automatically be symmetric.  His result turned out to be the Abraham tensor.  Gordon's  derivation has more recently been reconsidered by Antoci and Mihich\cite{Antoci}.  They again find the symmetric Abraham tensor describing the full system.

The momentum density in the Minkowski theory is ${\bf D}\times{\bf B}$ where ${\bf D}$ is the electric displacement field and ${\bf B}$ is the magnetic induction field.  This is the same as follows directly from the Maxwell equations in a medium. When the electromagnetic field is quantized, a free photon in the medium will then have a momentum $\bp = n\bp_0$ where $\bp_0$ is the momentum in vacuum. Since the momentum density in the Abraham theory is $n^2$ smaller, the corresponding photon will instead have the momentum $\bp = \bp_0/n$. It is strictly no longer a conserved quantity. 

A great deal of work has been done to experimentally confirm which description is correct. A detailed summary of both the theoretical and the experimental situation was given by Brevik\cite{Brevik}. Most experimental situations seem to be explained by the Minkowski tensor but there are a few important exceptions. In a more modern review it is concluded that there is no well-defined energy-momentum tensor for the electromagnetic field in a medium\cite{RMP}.  

In the following we will again consider the derivation of the energy-momentum tensor along the lines first suggested by Gordon\cite{Gordon}. For the field alone we find a new symmetric and conserved energy-momentum tensor with the same momentum density as in the Minkowski theory. On the other hand, for the full  system we recover the Abraham tensor. Due to the interaction with the medium, it is not conserved.  The results fit into the more phenomenological picture recently proposed by Barnett \cite{Barnett} and by Baxter and Loudon\cite{BL}.  The Minkowski momentum applies to single photons in the medium  while the Abraham momentum should be used for the electromagnetic field coupled to the medium.

Here we consider for simplicity a uniform, dielectric medium with the permittivity $\ve$. Using units so that the vacuum permittivity  $\ve_0$ and susceptibilty $\mu_0$ both have the values $\ve_0 = \mu_0 = 1$, the index of refraction is  $n = \sqrt{\ve}$. Effects of dispersion are not of interest for our discussion here. The Maxwell equations for the electric ${\bf E}$ and magnetic ${\bf B}$ fields are defined in the rest frame of the medium. Ignoring sources, they can be derived from the Lagrangian 
\beq
             {\cal L} =  {1\over 2} n^2{\bf E}^2 -  {1\over 2}{\bf B}^2                \label{EM-Lag}
\eeq
Introducing the vector potential ${\bf A}$, we have ${\bf E} = - \dot{\bf A}$ and ${\bf B} = {\bg\nabla}\times {\bf A}$. In this frame one can now introduce new four-vector coordinates $x^\mu = (t/n, \bx)$ and write the Lagrangian on a covariant form as in ordinary vacuum. The resulting energy-momentum tensor is conserved and symmetric. Its different components correspond to the energy and momentum densities of the Minkowski theory\cite{FR}.

Instead of introducing these new coordinates, we can alternatively introduce the tensor $\hat{\eta}^{\mu\nu} = \mbox{diag}(n^2, -1,-1,-1)$ which in some respects acts as a metric for the field in the medium. As first implicitly shown by Minkowski\cite{Minkowski} and later exploited by Gordon\cite{Gordon}, this point of view  allows for the description of the system when  it is in motion with constant velocity, i.e. implementation of  covariance under ordinary Lorentz transformations. When the medium has the constant three-velocity  ${\bf v}$,  the corresponding four-velocity is $u^\mu = \gamma(1, \bf{v})$ where $\gamma = (1 - v^2)^{-1/2}$ is the  Lorentz factor.  In such an inertial frame the above tensor becomes the more general Gordon metric ${\hat\eta}^{\mu\nu} = \eta^{\mu\nu} + (n^2 -1) u^\mu u^\nu$ where $\eta^{\mu\nu}$ is the contravariant Minkowski metric.  Notice that while the indices of any true four-vector like the four-velocity, are raised and lowered by the Minkowski metric, this is not the case with indices of the Gordon metric.

The derivation of the energy-momentum tensor is most simply demonstrated for a massless, scalar field in the medium. Instead of the Lagrangian (\ref{EM-Lag}), we will therefore consider the density
\beq
       {\cal L} = {1\over 2}n^2{\dot\phi}^2 -  {1\over 2}({\bg\nabla}\phi)^2           \label{S-Lag}
\eeq
when the field has the reduced propagation velocity $c = 1/n$ in the rest frame.  From the canonical field momentum $\Pi = \del{\cal L}/\del \dot{\phi} = n^2\dot{\phi}$ follows the Hamiltonian or equivalent energy density ${\cal H} =  (1/2)(n^2{\dot\phi}^2 +  ({\bg\nabla}\phi)^2)$. With the use of the  Gordon metric the Lagrangian can now be written as $ {\cal L} = (1/2)\hat{\eta}^{\mu\nu}\del_\mu\phi\del_\nu\phi$ in an arbitrary, inertial frame. The equation of motion for the field is then
 \beq
        {\hat\eta}^{\mu\nu}\del_\mu\del_\nu\phi =  [\del^\mu\del_\mu + (n^2-1)(u^\mu\del_\mu)^2]\phi = 0    \label{wave}
 \eeq
Needless to say, one will have a rather complex dispersion relation for the field excitations in such a frame.

Following  Gordon\cite{Gordon}, we  consider now the more general situation of the scalar field in a curved spacetime with the metric $g_{\mu\nu}$. Its action is then
\beq
            S[\phi] =  {1\over 2} \int\!d^4x \sqrt{-\hat{g}}\, \hat{g}^{\mu\nu}\del_\mu\phi\,\del_\nu\phi            \label{act}
\eeq
Here is  ${\hat g}^{\mu\nu} = g^{\mu\nu} + (n^2 -1) u^\mu u^\nu$ the corresponding Gordon metric and  the determinant $\hat{g} = \Det{(\hat{g}_{\mu\nu})}$. The medium four-velocity $u^\mu$ is taken to be constant. The contravariant components $T^{\mu\nu}$ of the energy-momentum tensor for the field can now be found by the variation $\hat{g}_{\mu\nu} \ra \hat{g}_{\mu\nu} + \delta\hat{g}_{\mu\nu}$ of the material Gordon metric. Here is $\hat{g}_{\mu\nu} = g_{\mu\nu} + (1/n^2 -1) u_\mu u_\nu$ the metric tensor inverse to ${\hat g}^{\mu\nu}$. Using the standard relations $\delta\hat{g}^{\mu\nu} = - \hat{g}^{\mu\alpha}\hat{g}^{\nu\beta}\delta\hat{g}_{\alpha\beta}$ and $\delta \sqrt{-\hat{g}} = (1/2)\sqrt{-\hat{g}} \hat{g}^{\alpha\beta}\delta\hat{g}_{\alpha\beta}$, we find the resulting variation of the action to be
\be
     \delta S[\phi] = - {1\over 2} \int\!d^4x \sqrt{-\hat{g}}\,[\hat{g}^{\alpha\mu}\hat{g}^{\beta\nu}\del_\mu\phi\,\del_\nu\phi                                                                                  - \hat{g}^{\alpha\beta}{\cal L}]\delta\hat{g}_{\alpha\beta}
\ee
Here ${\cal L} = (1/2)\hat{g}^{\mu\nu}\del_\mu\phi\del_\nu\phi$ is now the scalar Lagrangian density. In the original, flat spacetime we therefore have the energy-momentum tensor
\beq
            \hat{T}^{\mu\nu} =   \hat{\eta}^{\mu\alpha}\hat{\eta}^{\nu\beta}\del_\alpha\phi\,\del_\beta\phi - {1\over 2}\hat{\eta}^{\mu\nu}\hat{\eta}^{\alpha\beta}\del_\alpha\phi\,\del_\beta\phi           \label{SEM-tensor}
\eeq
It is obviously symmetric and easily seen to be conserved with use of the equation of motion (\ref{wave}).

In the rest frame of the medium we can now easily work out its different components. They can be assembled in the matrix
\beq
           \hat{T}^{\mu\nu} =  \left( \begin{array}{c | c} n^2{\cal H} & -n^2\dot{\phi}{\bg\nabla\phi} \\  \hline  -n^2\dot{\phi}{\bg\nabla\phi}  &\hat{T}_{ij}\end{array}\right)        
\eeq
where ${\cal H}$ is the previous rest-frame energy density and $\hat{T}_{ij}$ the spatial stresses for this scalar theory. We find the  energy current to be given by the vector ${\bf S} =  -\dot{\phi}{\bg\nabla\phi}$.  The factor $n^2$  is cancelled here and in $\hat{T}^{00}$ by the red shift $\hat{g}_{00} = 1/n^2$  when extracting the physical field energy content in the medium. This small complication is the prize to pay for using this metric formulation. However, this red-shift factor does not appear in the momentum components of the tensor. The momentum density is therefore given by the vector ${\bf G} = -n^2\dot{\phi}{\bg\nabla\phi}$.

Quantization of the scalar theory is straightforward in the rest frame.  The field operator can be expressed in terms of creation and annihilation operators with the canonical commutator $[a_\bk,a_{\bk'}^\dagger] = \delta_{\bk\bk'}$.  Integrating the  energy density ${\cal H}$, the full Hamiltonian takes the standard form $H = \sum_{\bk\lambda} \hbar\om_\bk \big(a_\bk^\dagger a_\bk  +1/2 \big)$ where  $\omega_\bk =  |\bk|/n$ as follows from the wave equation (\ref{wave}). Similarly,  the total field momentum ${\bf P} = \int\! d^3x\, {\bf G}$ becomes ${\bf P} = \sum_\bk \hbar\bk a_\bk^\dagger a_\bk$.  A field quantum with the wave number $\bk$ thus has the energy $\ve = \hbar\omega_\bk$ and the momentum $\bp = \hbar\bk$. Such quanta are well known in condensed matter physics.

So far we have only considered the energy-momentum tensor of the field. But it is locked to the medium and the full energy-momentum tensor ${T}^{\mu\nu}$ of the coupled system can be obtained from the variation $\delta{g}_{\mu\nu}$ of the full metric.  It can easily be found from the previous derivation\cite{Gordon}.  Introducing the vector $\hat{T}^\mu = \hat{T}^{\mu\nu}u_\nu$, the result can be written as 
\beq
          {T}^{\mu\nu} =  \hat{T}^{\mu\nu} + \Big({1\over n^2} - 1\Big)\Big(u^\mu \hat{T}^\nu  + u^\nu \hat{T}^\mu  - u^\mu u^\nu \hat{T}^\alpha u_\alpha\Big) \label{T-full}
\eeq
The second term represents the interaction with the medium and spoils the conservation of this energy-momentum tensor. But the two tensors are closely related. Defining ${T}^\mu = {T}^{\mu\nu}u_\nu$,  it then easily follows that $T^\mu = \hat{T}^\mu/n^2$.  

In this formalism we can now write the electromagnetic Lagrangian (\ref{EM-Lag}) as
\beq
          {\cal L} = - {1\over 4} F_{\mu\nu} H^{\mu\nu} 
\eeq
where $F_{\mu\nu} = \del_\mu A_\nu - \del_\nu A_\mu$ is the usual electromagnetic field tensor  with components in terms of ${\bf E}$ and ${\bf B}$. The tensor  $H^{\mu\nu} = \hat{\eta}^{\mu\alpha}\hat{\eta}^{\nu\beta} F_{\alpha\beta}$ contains the corresponding material fields ${\bf D}$ and ${\bf H}$.  In the frame where the dielectric under consideration is at rest,  ${\bf D} = n^2 {\bf E}$ and ${\bf H} =  {\bf B}$. Our treatment can easily be made  valid for the more general situation with a magnetic permeability $\mu \neq 1$ as considered by Minkowski and Gordon.

The wave equation in the absence of free charges and currents will now follow from $\del_\mu H^{\mu\nu} = 0$. The two other Maxwell equations follow from the ordinary Bianchi identity $\del_\lambda F_{\mu\nu} + \del_\nu F_{\lambda\mu}  + \del_\mu F_{\nu\lambda} = 0$. Going through the same steps as for the scalar case, we obtain the following energy-momentum tensor for the electromagnetic field
\beq
          \hat{T}^{\mu\nu} = \hat{\eta}^{\mu\alpha} F_{\alpha\beta} H^{\beta\nu} -  \hat{\eta}^{\mu\nu} {\cal L}
\eeq
It has a structure very similar to the Minkowski energy-momentum tensor, but is found to be both symmetric and conserved with the use of the equations of motion. 

In the rest frame of the medium it has the components
\beq
   \hat{T}^{\mu\nu} =  \left(\begin{array}{c | c} n^2{\cal H} & {\bf D}\times{\bf B} \\  \hline  {\bf D}\times{\bf B}  &\hat{T}_{ij}\end{array}\right)      \label{M-EM}  
\eeq
where ${\cal H}$ is the ordinary electromagnetic energy density and $\hat{T}_{ij}$ are the Maxwell stresses. Since ${\bf D}\times{\bf B} = n^2 {\bf E}\times{\bf H}$ we recover the standard Poynting vector ${\bf S} = {\bf E}\times{\bf H}$ for the energy current when we invoke the same red-shift  factor as in the scalar case. The momentum density is ${\bf G} = {\bf D}\times{\bf B}$ as also results from the  Minkowski theory. Quantization can be performed in the rest frame as for the scalar field resulting in photons with the dispersion relation  $\ve = p/n$. 

When we now turn to the the energy-momentum tensor of the full system  (\ref{T-full}), it turns out to be exactly the Abraham tensor. This is most easily seen in the rest system of the medium where $\hat{T}^\mu = \hat{T}^{\mu 0}$. Thus we have
\beq
      {T}^{\mu\nu} =  \left(\begin{array}{c | c} {\cal H} & {\bf E}\times{\bf H} \\  \hline  {\bf E}\times{\bf H}  &\hat{T}_{ij}\end{array}\right)      \label{A-EM}  
\eeq
In this case there is no red-shift factor since $g_{00} = 1$.

The consistent use of the Gordon metric  enables us to derive a symmetric and conserved energy-momentum tensor for the electromagnetic field in a moving medium. It has a physical content which is very similar to what is contained in the original  Minkowski tensor. The unwanted asymmetry of the latter is in our derivation replaced by a red shift which is direct consequence of the Gordon metric in the medium. For other field excitations in continuous media like sound waves, one should expect that the effects of these two tensors should manifest themselves in similar ways as in the electromagnetic case.

We want to thank Cliff Burgess for several clarifying discussions and Konrad Tywoniuk for a very useful suggestion.


\begin{thebibliography}{99}

\bibitem{Minkowski} H. Minkowski, {\it Nachr. Ges. Wiss. G\"ottingen}, 53 (1908); {\it Math. Ann.} {\bf 68}, 472 (1910).

\bibitem{Abraham} M.  Abraham,  {\it Rend. Circ. Mat. Palermo} {\bf 28}, 1 (1909); {\it ibid.} {\bf 30}, 5 (1910).

\bibitem{Gordon} W. Gordon,  {\it Ann. d. Phys.} {\bf 72},  421 (1923).

\bibitem{Antoci} S. Antoci and L. Mihich, {\it Nuovo Cimento} {\bf B112}, 991 (1997); gr-qc/970455.

\bibitem{Brevik}  I. Brevik,  {\it Phys. Rep.}  {\bf 52}, 133 (1979).

\bibitem{RMP} R. N. C. Pfeifer, T. A. Nieminen, N. R. Heckenberg and H. Rubinsztein-Dunlop, {\it Rev. Mod. Phys.} {\bf 79}, 1197 (2007).

\bibitem{Barnett}  S.M. Barnett, {\it Phys. Rev. Lett.} {\bf 104}, 070401 (2010).

\bibitem{BL} C. Baxter and R. Loudon, {\it Journal of Modern Optics} {\bf 57},  830 (2010).

\bibitem{FR} F. Ravndal, arXiv:0804.4013[quant-ph], arXiv:0810.1872[hep-ph].

\end{thebibliography}
\end{document}